# Charge-state distributions of $^{20}$Ne ions emerging from thin foils

M. Cavallaro[1*], G. Santagati[1], F. Cappuzzello[1,2], D. Carbone[1], R. Linares[3], D. Torresi[1], L. Acosta[4], C. Agodi[1], D. Bonanno[5], D. Bongiovanni[1], T. Borello-Lewin[6], I. Boztosun[7], S. Calabrese[1,2], D. Calvo[8], E.R. Chavez Lomeli[4], P.N. De Faria[3], F. Delaunay[8,9,11], N. Deshmukh[1], P. Finocchiaro[1], M. Fisichella[8], A. Foti[5], G. Gallo[1,2], A. Hacisalihoglu[1,10], F. Iazzi[8,11], R. Introzzi[8,11], G. Lanzalone[1,12], D. Lo Presti[2,5], F. Longhitano[5], N.H. Medina[6], A. Muoio[1], J.R.B. Oliveira[6], A. Pakou[13], L. Pandola[1], H. Petrascu[14], F. Pinna[8,11], S. Reito[5], G. Russo[2,5], O. Sgouros[1,13], S.O. Solakci[7], V. Soukeras[1,13], G. Souliotis[15], A. Spatafora[1,2], S. Tudisco[1], A. Yildirim[7], V.A.B. Zagatto[3]

for the NUMEN collaboration

[1] Istituto Nazionale di Fisica Nucleare, Laboratori Nazionali del Sud, Catania, Italy

[2] Dipartimento di Fisica e Astronomia "Ettore Majorana", Università degli Studi di Catania, Catania, Italy

[3] Instituto de Física, Universidade Federal Fluminense, Niterói, Brazil

[4] Instituto de Física, Universidad Nacional Autónoma de México. Apartado Postal 20-364, Cd. Mx. 01000, Mexico

[5] Istituto Nazionale di Fisica Nucleare, Sezione di Catania, Catania, Italy

[6] Instituto de Física, Universidade de São Paulo, São Paulo, Brazil

[7] Akdeniz University, Antalya, Turkey

[8] Istituto Nazionale di Fisica Nucleare, Sezione di Torino, Torino, Italy

[9] LPC Caen, Normandie Université, ENSICAEN, UNICAEN, CNRS/IN2P3, Caen, France

[10] Institute of Natural Science, Karadeniz Teknik Universitesi, Trabzon, Turkey

[11] DISAT-Politecnico di Torino, Torino, Italy

[12] Università degli Studi di Enna "Kore", Enna, Italy

[13] Department of Physics and HINP, The University of Ioannina, Ioannina, Greece

[14] IFIN-HH, Bucharest, Romania

[15] Laboratory of Physical Chemistry, Department of Chemistry, National and Kapodistrian University of Athens, Athens

*manuela.cavallaro@lns.infn.it

**Abstract**

New experimental measurements of charge state distributions produced by a $^{20}$Ne$^{10+}$ beam at 15 MeV/u colliding on various thin solid targets are presented. The use of the MAGNEX magnetic spectrometer enabled measurements of the 8$^+$ charge state down to fractions of a few 10$^{-5}$. The use of different post-stripper foils located downstream of the main target is explored, showing that low *Z* materials are particularly effective to shift the charge state distributions towards fully stripped conditions. The dependence on the foil thickness is also studied and discussed.

1.         **Introduction**

Magnetic devices, such as spectrometers or beam analyzers, are usually employed in high-precision modern nuclear physics experiments. In such measurements, the different ionic charge states produced by the interaction of the accelerated ion beam with thin foils used as degraders or as nuclear targets can constitute alternatively a source of information or an unwanted background.

A case where an accurate and precise study of charge-state distributions is crucial is the measurement at very forward angles of the products of nuclear reactions induced by medium-heavy ions (A > 4) at energies above the Coulomb barrier. The NUMEN (Nuclear Matrix Elements for Neutrinoless double beta decay) project [1], [2] aims at investigating the nuclear response to double charge exchange (DCE) reactions for all the isotopes of interest for neutrinoless double beta decay (0νββ) at INFN – Laboratori Nazionali del Sud (Italy). Typical used beams are $^{18}$O$^{8+}$, $^{20}$Ne$^{10+}$, $^{12}$C$^{6+}$ at energies between 15 and 70 MeV/u and typical targets are nuclei candidates for 0νββ, i.e. specific medium-mass nuclei such as $^{48}$Ca, $^{76}$Ge, $^{116}$Cd, $^{130}$Te and others. The reaction ejectiles are momentum analyzed by the MAGNEX large acceptance magnetic spectrometer [3] and detected by its focal plane detector [4], [5].

The DCE reaction channel is characterized by very low cross sections in the ground to ground state transitions (from few μbarn down to few nbarn) [6], [7]. On the other hand, the elastically and inelastically scattered ions with different charge states produced at the target and having the same magnetic rigidity of the reaction products of interest, can reach the focal plane detector at very high rate, especially at forward angles [8]. For example, the $^{20}$Ne$^{10+}$ beam impinging on a target produces, among other ejectiles, $^{20}$O$^{8+}$ (DCE reaction) and $^{20}$Ne$^{8+}$ (elastic scattering). Both have almost the same magnetic rigidity and the yields of the $^{20}$Ne$^{8+}$ significantly interfere with the $^{20}$O$^{8+}$ detection. The use of high intensity incident beams, desired to face the low cross-sections of the DCE processes of interest, is consequently prevented by the intolerable overall count rate at the detectors [9].

In these experiments, the choice of the target material is imposed by the physics requirements. A possible solution is to place a secondary foil, located between the target and the magnetic elements, working as a stripper (herein referred to as post-stripper), to re-distribute the exiting beam charge states reducing the amount of unwanted lower charge states. The aim is to find the best conditions that minimize the contribution of the unwanted charge state scattering processes and allow to increase the beam intensity and consequently the overall experimental sensitivity to cross section measurements. Therefore, an accurate knowledge of the charge state distribution $F(q)$ emerging from different foils is crucial.

The study of $F(q)$ for different beam/target combination is a subject of large interest since the '40s. The first experimental investigation from uranium fission fragments go back to the work of Lassen [10]. He first observed a density effect: higher degrees of ionization are obtained when ions emerge from solid rather than gases or when the pressure of the gas is increased.

Slightly higher velocities of the incident ions started to be investigated with the advent and development of accelerators for heavy ions in the 1970s and 1980s, with the main purpose to properly design and optimize such machines.

Due to the limited comprehensive theories, often a phenomenological approach has been pursued, leading to empirical formulas for the mean charge ($\bar{q}$), width ($d$) or shape of the equilibrium charge distribution, based on the available data. One should cite, for example, the semi-empirical method of Dmitriev and Nikolaev [11], which is an improvement of the Bohr generalized criterion for the conditions of the loss and retention of electrons by atomic particles passing through matter, or the report of Betz [12], that summarizes the main concepts in the field.

These and other works of that period [13], [14], [15], [16] typically refer to incident energies below 2 MeV/u and ions which still carry many electrons in their atomic shells [12]. When $F(q)$ is dominated by fully stripped ions ($q = Z$), the results of empirical formulas for $\bar{q}$ and $d$, valid at low velocities [11], [17], [18], diverge systematically with increasing velocity and $Z$ and the Gaussian distribution is no longer an appropriate expression for $F(q)$ [19].

In the 1990s an important reanalysis of the available data was done [19], producing tables for the charge distribution of few-electron ions with $q = Z$, $Z-1$ and $Z-2$ and for ion energies up to 40 MeV/u. These tables contain evaluated data obtained by interpolating existing experimental data. Moreover, they regard only specific cases and only the passage of ions through carbon foils. The behavior of charge state distributions for mediums other than carbon is much less explored in the literature [18], [20].

The thickness of the target material at which a charge equilibration occurs has also represented an issue in past studies. When energetic ions pass through a medium, their charge states vary as a function of the penetration depth. At a certain depth, charge equilibration is attained and the charge distributions become independent of the initial charge state of the ions [21]. Betz [22], Baron [14] and Zaikov et al. [23] have attempted to establish some trend of the relationship between charge equilibrium foil thickness and ion atomic number and kinetic energy. According to Ref. [21], after charge equilibration is achieved, the charge distribution nevertheless varies as a function of penetration depth. The variation of $F(q)$ and $\bar{q}$ in the charge equilibration regime of foil thicknesses is found to be connected to the energy loss in traversing the foil. However the existing experimental data are again limited to low energies (1-4 MeV/u, maximum 1 MeV/u in the case of Ne beam), and to ions with many electrons in the atomic shells. Using empirical relations connecting $\bar{q}$ and foil thickness (or emergent energy) [11], a disagreement is found in the absolute value [21], while attempts to fit the stripper equilibrium thickness versus beam energy by analytical functions are not successful especially at energies higher than 10 MeV/u, where experimental data were missing [14]. So an absolute and general way for the determination of equilibrium thickness is presently not available.

In this work, we present new experimental data for a quantitative study of the charge state distribution generated by a $^{20}$Ne$^{10+}$ beam at 15 MeV/u on Au, Cd, Ge and C targets coupled with different post-stripper materials and thickness.

## 2. The measurements

Several experimental tests have been performed at INFN - Laboratori Nazionali del Sud (INFN-LNS) using $^{20}$Ne beams accelerated by the K800 Superconducting Cyclotron at 15 MeV/u. In each experiment, the beam impinges on a different target located in a scattering chamber. The targets are constituted by a main thin foil (made of the isotopic material of interest for the nuclear physics studies) followed by a post-stripper foil. The main and post-stripper foils used in the experiments and their thickness are listed in Table 1. The Au targets are self-supporting and produced by evaporation, the Te and Ge targets are evaporated on a thin (~30 µg/cm$^2$) carbon foil while the Cd targets are produced by rolling. They are manufatured at the chemical laboratory of INFN-LNS. The target thickness is determined by measuring the energy loss in the target of α-particles from a collimated radioactive source. An estimated uncertainty of ±10% affects the thickness measurements.

The elastically scattered ions emerging from the target at forward angles with different charge states $q = Z$ ($^{20}$Ne$^{10+}$), $Z - 1$ ($^{20}$Ne$^{9+}$), $Z - 2$ ($^{20}$Ne$^{8+}$) are momentum analyzed by the MAGNEX large acceptance magnetic spectrometer [3], [24] positioned with its optical axis centered at $\theta_{opt} = 8°$. Thanks to the large angular aperture of the spectrometer, this setup corresponds to a measured angular range $3° < \theta < 14°$. The choice of the angle was done to maximize the count rate at the detector;

however, we do not observe any angular dependence of the measured charge state distribution in the explored region. All the results shown here are integrated within the measured angular range. The ejectiles are identified in mass number *A*, atomic number *Z* and charge state *q* following the technique described in Ref. [25].

For each target/post-stripper configuration, three runs are performed at different magnetic settings of the spectrometer, each one focusing at the focal plane different $^{20}$Ne charge state ejectiles with $q = Z$, $q = Z - 1$ and $q = Z - 2$. For each run, the total charge is integrated by a Faraday cup downstream of the target. The yields normalized to the total integrated charge and corrected for the acquisition dead time and the detector efficiency [25], [26] are extracted.

Limiting the treatment of the charge distribution to three components with $q = Z$, $q = Z - 1$, $q = Z - 2$, the fractions of ions $F(q)$ for a given charge state $q$ is derived from the experimental data and is shown in Table 1 and graphically represented in Figs. 1 and 2.

The error in the determination of $F(q)$ is evaluated taking into account (i) the statistical uncertainty in the measurement of the yields (maximum estimated ~3% in the 8$^+$ runs, which are characterized by low count rate), (ii) the Faraday Cup accuracy (maximum estimated ~1%, mainly due to the charge collection measurement in the 10$^+$ runs characterized by low beam intensity) and (iii) the spectrometer acceptance (maximum ~2%). Other sources of systematic error in the Faraday cup measurement are cancelled in the ratio.

Table 1 List of the target/post-stripper configurations used in the experimental runs. The charge distribution $F(q)$ for $q = Z$, $q = Z - 1$, $q = Z - 2$, the average charge $\bar{q}$ and the charge distribution width *d* are extracted as described in the text.

| Target material | Target thickness [μg/cm$^2$] | Post-stripper material | Post-stripper thickness [μg/cm$^2$] | F(8$^+$) | F(9$^+$) | F(10$^+$) | $\bar{q}$ | *d* |
|---|---|---|---|---|---|---|---|---|
| $^{197}$Au | 940 | no post-stripper | - | 5.0E-03 | 1.4E-01 | 0.85 | 9.85 | 0.372 |
| $^{197}$Au | 920 | $^{27}$Al | 1080 | 7.7E-04 | 9.0E-02 | 0.91 | 9.91 | 0.290 |
| $^{197}$Au | 420 | LiF + C | 980 | 2.3E-04 | 4.1E-02 | 0.96 | 9.96 | 0.158 |
| $^{197}$Au | 930 | C$_{10}$H$_8$O$_4$ (Mylar) | 840 | 9.2E-05 | 2.7E-02 | 0.97 | 9.97 | 0.164 |
| $^{197}$Au | 980 | C | 990 | 7.7E-05 | 2.5E-02 | 0.97 | 9.97 | 0.158 |
| $^{197}$Au | 990 | C$_2$H$_4$ (HDPE) | 950 | 3.5E-05 | 1.5E-02 | 0.98 | 9.99 | 0.122 |
| $^{197}$Au | 920 | $^9$Be | 1000 | 3.2E-05 | 7.3E-03 | 0.99 | 9.99 | 0.086 |
| $^{116}$Cd | 1080 | no post-stripper | - | 4.5E-03 | 5.1E-02 | 0.94 | 9.94 | 0.257 |
| $^{116}$Cd | 1330 | C$_2$H$_4$ (HDPE) | 950 | 5.2E-05 | 8.7E-03 | 0.99 | 9.99 | 0.094 |
| $^{116}$Cd | 1360 | C$_3$H$_6$ (PP) | 360 | 4.8E-05 | 8.2E-03 | 0.99 | 9.99 | 0.091 |
| $^{116}$Cd | 1370 | C$_3$H$_6$ (PP) | 720 | 5.8E-05 | 8.3E-03 | 0.99 | 9.99 | 0.092 |
| $^{116}$Cd | 1080 | C$_3$H$_6$ (PP) | 1080 | 6.9E-05 | 1.1E-02 | 0.99 | 9.99 | 0.106 |
| $^{130}$Te | 250 | no post-stripper | - | 3.9E-04 | 4.3E-02 | 0.96 | 9.96 | 0.206 |
| $^{130}$Te | 250 | C$_2$H$_4$ (HDPE) | 950 | 8.4E-05 | 1.2E-02 | 0.99 | 9.99 | 0.111 |
| $^{130}$Te | 250 | C$_3$H$_6$ (PP) | 360 | 3.8E-05 | 8.2E-03 | 0.99 | 9.99 | 0.091 |
| $^{130}$Te | 240 | C$_3$H$_6$ (PP) | 720 | 3.8E-05 | 9.8E-03 | 0.99 | 9.99 | 0.100 |
| $^{130}$Te | 240 | C$_3$H$_6$ (PP) | 1080 | 6.9E-05 | 1.1E-02 | 0.99 | 9.99 | 0.106 |
| $^{130}$Te | 250 | C | 920 | 1.0E-04 | 1.3E-02 | 0.99 | 9.99 | 0.117 |
| $^{76}$Ge | 390 | C | 300 | 5.2E-05 | 2.0E-02 | 0.98 | 9.98 | 0.139 |
| $^{76}$Ge | 400 | C | 800 | 9.1E-05 | 2.8E-02 | 0.98 | 9.97 | 0.167 |
| C | 900 | - | - | 9.8E-05 | 1.6E-02 | 0.98 | 9.98 | 0.122 |
| C Shima | not available | - | - | 2.0E-05 | 8.8E-03 | 0.99 | 9.99 | 0.091 |

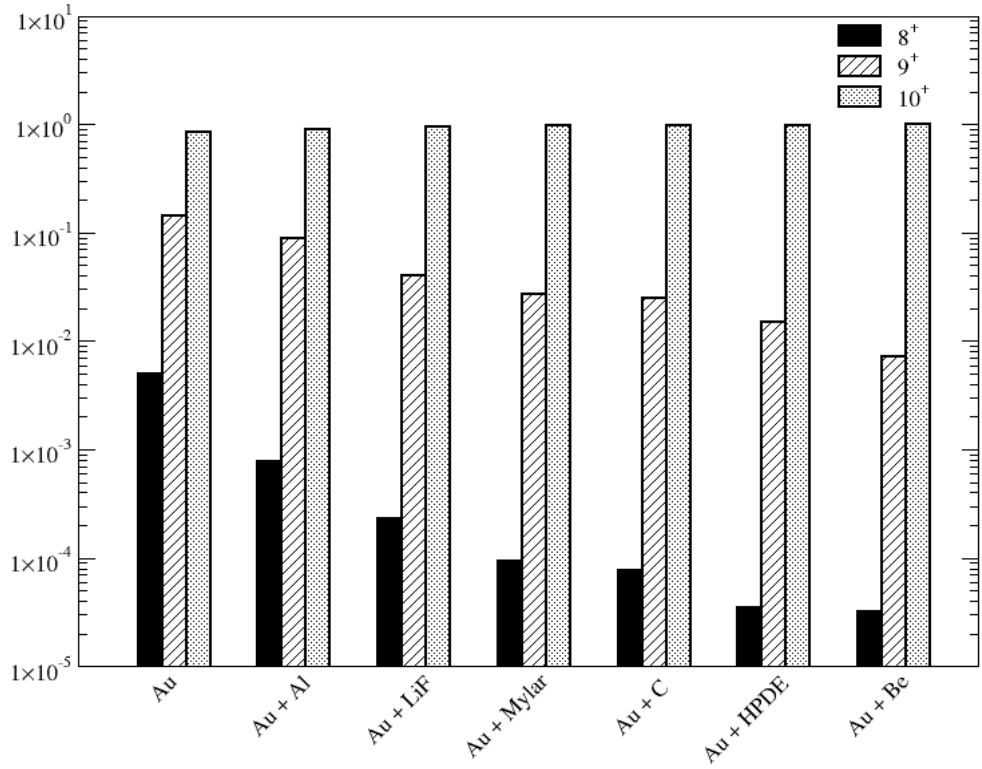

Fig. 1. Plot of the measured charge state distribution generated by a $^{20}$Ne$^{10+}$ beam at 15 MeV/u exiting pure Au target and Au targets followed by different post-stripper materials. Details are listed in Table 1.

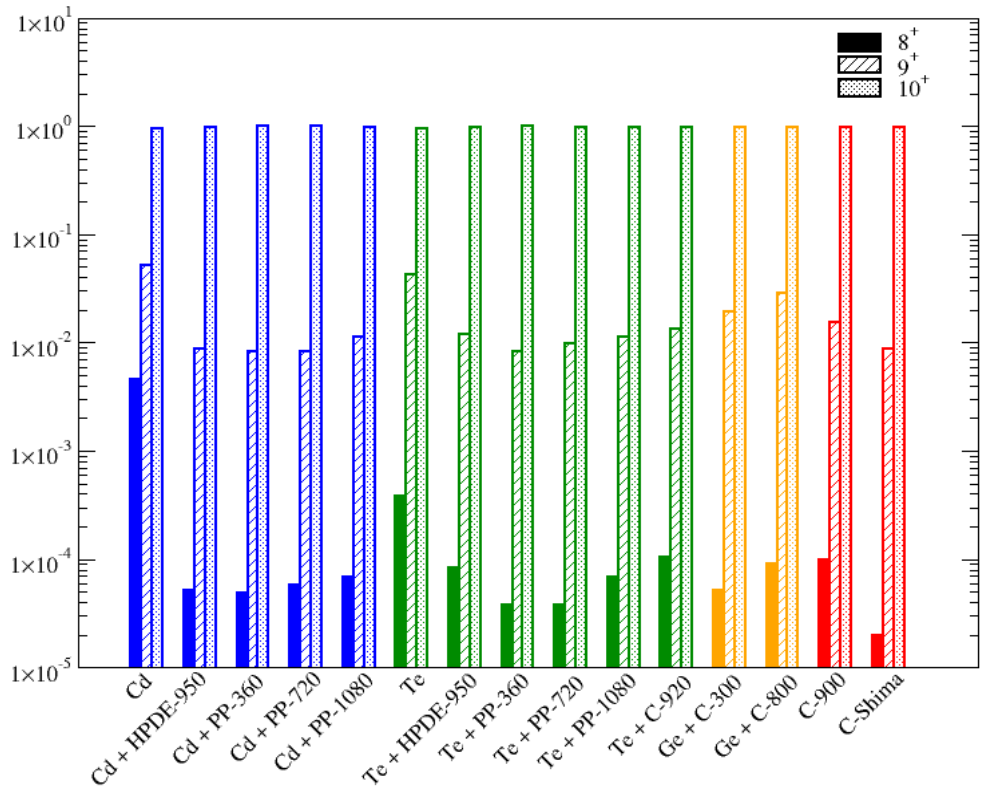

Fig. 2. Plot of the measured charge state distribution produced by a $^{20}$Ne$^{10+}$ beam at 15 MeV/u exiting Cd (blue), Te (green), Ge (orange) and C (red) targets without and with post-strippers. Details are listed in Table 1.

## 3. Study of different post-stripper foils

The mean charge $\bar{q}$ and the charge distribution width (standard deviation) $d$ is defined as:

$$\bar{q} = \sum_q q F(q) \quad (1)$$

$$d^2 = \sum_q (q - \bar{q})^2 F(q) \quad (2)$$

Fig. 3 shows the plot of the charge distribution width $d$ against the mean number of electrons remaining in the ion $\overline{n_e} = Z - \bar{q}$. For all the systems studied here, the relation between $d$ and $\bar{q}$ is in agreement with the universal behavior $d = \overline{n_e}^{1/2}$ observed in Ref. [19] for several beams impinging on carbon target at energies lower than ours. This relation follows from eqs. (1) and (2) when the charge distribution is dominated by fully stripped ions ($q = Z$). Our result is thus a consequence of the small percentage of $q = Z - 1$ and $q = Z - 2$ in the present experimental conditions.

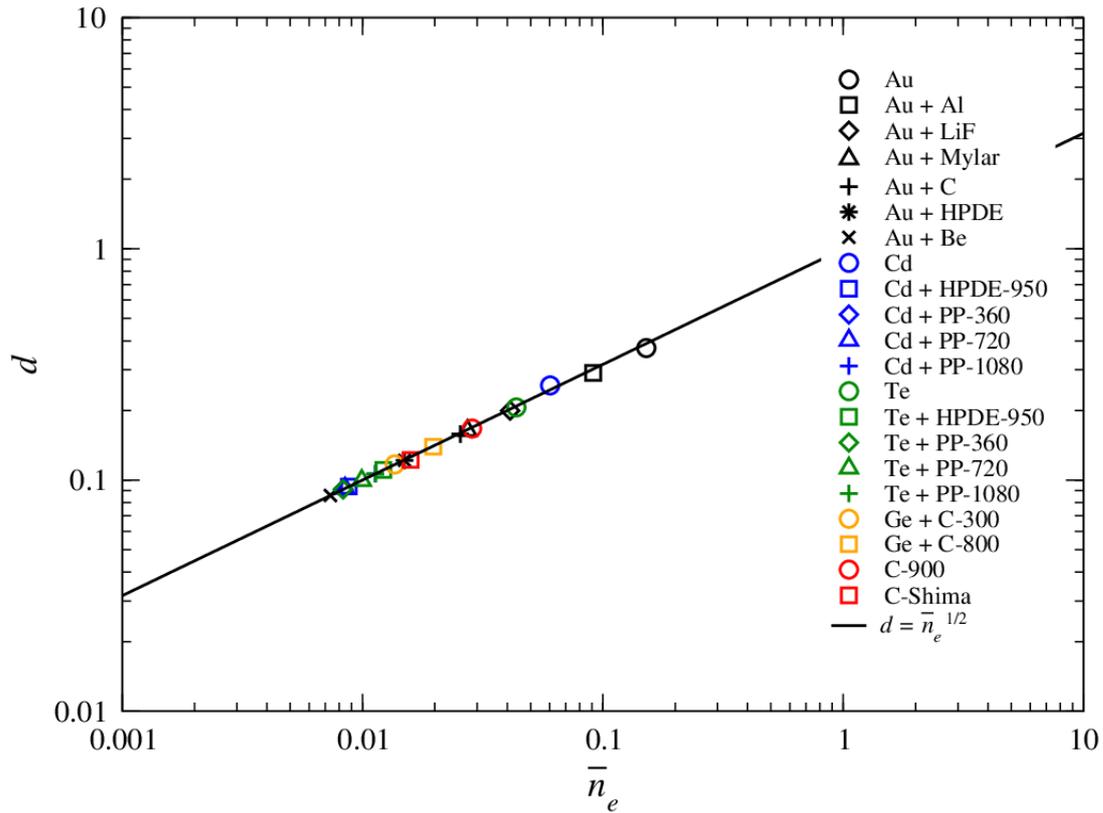

Fig. 3. Plot of the charge distribution width $d$ of Ne ions measured in the present experiments as a function of the mean number of electrons remaining in the ion $\overline{n_e} = Z - \bar{q}$. The line $d = \overline{n_e}^{1/2}$ is related to the charge distribution composed of only two charge states with $q = Z$ and $q = Z-1$.

In the first test, a fully stripped $^{20}$Ne$^{10+}$ beam at energy 15 MeV/u bombarded a number of target systems constituted by a $^{197}$Au foil followed by different post-stripper materials. All the investigated target/post-stripper configurations with the corresponding thicknesses are listed in Table 1. The resulting $F(q)$ are also shown in Table 1 and graphically represented in Fig. 1.

We note that for the no-post-stripper case, the presence of the sole Au target generates a re-distribution of the beam charge states from fully stripped $10^+$ to an average charge $\bar{q} = 9.85$ with charge distribution width $d = 0.372$.

The effect of the post-stripper material is evident. The fraction of $9^+$ decreases by more than one order of magnitude, while the $8^+$ is reduced by more than two order of magnitudes in the best cases. A clear trend for the stripping efficiency depending on the atomic number $Z$ of the post-stripper material is also observed. The average charge state $\bar{q}$ increases moving from Al to Be and the fraction of lower charge states decreases. Materials with lower average $Z$ are more efficient to maintain the beam fully stripped and to reduce the amount of $9^+$ and $8^+$ components. The carbon based materials, i.e. Mylar, C, polypropylene (PP), and high-density polyethylene (HDPE) and the beryllium foil seem the best candidates to minimize the lower charge state components.

Taking into account that the use of beryllium foils has some practical drawback due to its poisonousness, we have chosen to explore in more details the behavior of the carbon-based materials. In another experimental campaign, the $^{20}$Ne$^{10+}$ beam at 15 MeV/u energy was used to bombard targets that are of interest for the 0νββ research, namely $^{76}$Ge, $^{116}$Cd and $^{130}$Te. Each target was followed by a different carbon-based post-stripper material, as shown in Table 1. The obtained charge state fractions are listed in Table 1 and plotted in Fig. 2. Once again, the effect of the presence of a post-stripper material is important leading to a significant increase of the $10^+$ components.

In particular, we have studied the dependence of the charge state distribution on the post-stripper thickness. Thick post-strippers give slightly worse performances than thin ones, resulting in more $9^+$ and $8^+$ components. For example, in the case of $^{116}$Cd and $^{130}$Te targets followed by PP post-strippers, the 360 and 720 μg/cm$^2$ thicknesses seem to be equivalent, whereas in the case of 1080 μg/cm$^2$ thickness, the $\bar{q}$ is smaller. The same occurs in the case of $^{76}$Ge target followed by C post-stripper of two different thicknesses: the thin carbon (300 μg/cm$^2$) produces less $9^+$ and $8^+$ than the thick one (800 μg/cm$^2$). The reason is probably due to the fact that thicker foils reduce the emergent kinetic energy, shifting the average charge state toward lower values, as qualitatively described in Ref. [21] for 65 MeV Cu$^{9+}$ ions on C target. However, in our case the situation is more complex as we are studying a system with two foils (target and post-stripper). The effect of the contact between such foils could in fact play a role, deserving further analysis.

The effects produced by a single C stripper have been compared with the tabulated values reported in Ref. [19]. In our experimental data a 900 μg/cm$^2$ thick foil has been used. The thickness of the carbon foil used in Ref. [19] is not mentioned in the paper. The measured $F(q)$ and $\bar{q}$ are listed in Table 1 and represented in Fig. 2. In our data, the lower charge state components are more populated than in the literature data. This is likely a consequence of the larger thickness of our C foil.

### 4. Conclusions

In this paper, new experimental measurements of charge state distributions produced by a $^{20}$Ne$^{10+}$ beam at 15 MeV/u colliding on various thin solid targets followed by different secondary foils have been presented.

As expected, the beam remains almost fully stripped (10$^+$) after the interaction with the target. However, the lowest emerging charge states (9$^+$ and 8$^+$) are not negligible for nuclear physics experiments which use magnetic spectrometers. The experimental facility used here allowed us to measure even fractions of the order of 10$^{-5}$ of 8$^+$ charge state contributions. The effect of different post-stripper foils located downstream of the main target have been explored, showing that low *Z* materials are particularly effective to shift the charge state distributions towards fully stripped conditions. The use of specific target/post-stripper configurations is concluded to be very useful to improve the sensitivity of cross section measurement with magnetic spectrometers.

**Acknowledgments**

The authors thank C. Marchetta and A. Massara of the INFN-LNS chemical laboratory and all the INFN-LNS staff involved in the experimental activity described in the paper. This project has received funding from the European Research Council (ERC) under the European Union's Horizon 2020 research and innovation programme (grant agreement No 714625)**.** The projects PAPIIT IG101016 IA103218, PIIF2018 and CONACyT 294537 are also acknowledged.